\documentclass[12pt, a4paper]{article}

\usepackage[utf8]{inputenc}
\usepackage[left=2cm,right=2cm,top=2cm,bottom=3cm]{geometry}
\usepackage{amsmath}
\usepackage{mathrsfs}
\usepackage{amsfonts}
\usepackage{graphicx}
\usepackage{authblk}
\usepackage{subcaption}
\usepackage{csquotes}
\usepackage{mathtools}
\usepackage{xcolor}
\usepackage[T1]{fontenc}
\usepackage{hyperref}
\hypersetup{
    colorlinks=true,
    linkcolor=blue,
    filecolor=magenta,      
    urlcolor=cyan,
    pdftitle={Overleaf Example},
    pdfpagemode=FullScreen,
    }
\usepackage{tikz}
\usetikzlibrary{positioning}
\usetikzlibrary{matrix,arrows}
\usetikzlibrary{calc}
\usetikzlibrary{shapes, shapes.geometric, shapes.symbols, shapes.arrows, shapes.multipart, shapes.callouts, shapes.misc,decorations.pathmorphing}
\tikzset{snake it/.style={decorate, decoration=snake}}
\usepackage{tikz-3dplot}

\providecommand{\keywords}[1]
{
  \small	
  \textbf{\textit{Keywords---}} #1
}

\newtheorem{defn}{Definition}

\newcommand{\ri}{\rightarrow}

\title{
Violating the KCBS inequality with a toy mechanism
}
\author{Alisson Tezzin\footnote{alisson.tezzin@usp.br}}
\affil{\small\textit{Department of Mathematical Physics, Institute of Physics,
University of São Paulo
\\
R. do Matão 1371, São Paulo 05508-090, SP, Brazil}}
\date{\today}

\begin{document}

    \maketitle
    \begin{abstract}
    In recent years, much research has been devoted to exploring contextuality in systems that are not strictly quantum, like classical light, and many theory-independent frameworks for contextuality analysis have been developed. It has raised the debate on the meaning of contextuality outside the quantum realm, and also on whether --- and, if so, when --- it can be regarded as a signature of non-classicality. In this paper, we try to contribute to this debate by showing a very simple ``thought experiment'' or ``toy mechanism'' where a classical object (i.e., an object obeying the laws of classical physics) is used to generate experimental data violating the KCBS inequality. As with most thought experiments, the idea is to simplify the discussion and to shed light on issues that in real experiments, or from a purely theoretical perspective, may be cumbersome. We give special attention to the distinction between classical realism and classicality, and to the contrast between contextuality within and beyond quantum theory.
    \end{abstract}
    

\section{Introduction}
Kochen-Specker contextuality (in short, contextuality) is currently understood as the obstruction to the existence of a global probability distribution accounting for joint measurements performed on a physical system \cite{Abramsky_2011, budroni2021quantum}, and it has been shown that the emergence of contextuality in experimental data is of theoretical and practical importance \cite{budroni2021quantum,  howard2014contextuality, Gupta2022}. This notion of contextuality is ``theory-independent'', which means that it does not depend on any specific physical theory \cite{NaturePopescu, amaral2018graph}, and many theory-independent frameworks to contextuality have been proposed in recent years \cite{Abramsky_2011, CSW, budroni2021quantum, amaral2018graph}. This is also an ``operational'' or ``empiricist'' view on contextuality, given that it is experimental data that is labeled as ``contextual'' or ``noncontextual'' \cite{budroni2021quantum, deRondeOmelette, amaral2018graph}. 

In the particular case of Bell scenarios, i.e., scenarios representing freely chosen measurements performed in spacelike separated parts of a system, it is widely accepted that, regardless of what is meant by ``measurement'' or even by ``physical system'' \cite{NaturePopescu}, theory-independent contextuality, which in this case consists in the violation of  Bell inequalities \cite{amaral2018graph}, is only possible if non-local phenomena occur \cite{amaral2018graph, BrunnerBell, NaturePopescu}. And since non-locality  is a non-classical concept, theory-independent contextuality in Bell scenarios is broadly accepted  as a signature of non-classicality, i.e., it is a criterion telling us when, as Michael D. Mazurek et al. put it, ``nature fails to respect classical physics'' \cite{NatureMazurek}. Without spacelike separation, on the other hand, things are different. In fact, though theory-independent contextuality (by which we mean, from now on, contextuality without spacelike separation) is of undeniable importance \cite{barbosa2021closing, AmaralWirings}, going as far as to be studied outside the realm of physics \cite{Matt2019, cervantes2018snow, wang2021analysing}, there is no physical phenomenon, like non-locality, that is necessarily associated with it, let alone a non-classical phenomenon, thus the phenomenological significance of contextuality is a matter of debate \cite{deRondeOmelette, MarianRule, SvozilClicks}. Furthermore, the meaning of contextuality in quantum theory strongly depends on the Kochen-Specker theorem \cite{sep-kochen-specker, Kochen1975}, which basically asserts that one cannot consistently interpret all selfadjoint operators of a quantum system as representing properties simultaneously possessed by it \cite{Kochen1975, WhatThing}. Kochen-Specker theorem is essentially about classical \textit{realist} descriptions of quantum systems \cite{Kochen1975, WhatThing}, and the idea of realism is unsuitable for most modern theory-independent approaches to physics. Hence, the distinction between contextuality within and beyond quantum theory is a relevant one. In quantum theory, though some authors dispute it \cite{deRondeOmelette, SvozilClicks, sep-kochen-specker}, the view that contextuality is a signature of non-classicality is widespread  \cite{budroni2021quantum, Gupta2022, barbosa2021closing, sep-kochen-specker}. In theory-independent contextuality, on the other hand, it is easier to find both views in the literature \cite{SvozilClicks, barbosa2021closing, li2017experimental, Leo, ZhangExperimental, MarianRule}.

For all these reasons, the question of whether contextuality is a signature of non-classicality  --- and, if so, under which conditions and in which sense --- is relevant and much research, both theoretical and experimental, has been devoted to it \cite{deRondeOmelette, MarianRule, li2017experimental, li2019state, ZhangExperimental, Gupta2022}. In this paper, we try to contribute to this debate by showing a quite simple ``thought experiment'' or  ``toy mechanism'' where a classical object (i.e., an object that obeys the laws of classical physics) generates contextual data. Thought experiments are helpful tools to clarify difficult questions, and we believe that the one we propose here can shed some light on the conditions an experimental test of contextuality must satisfy in order to have phenomenological significance, i.e., in order to allow us to conclude that a certain kind of physical phenomenon has occurred. More importantly, it can help us to understand in which sense contextuality can be seen as a signature of non-classicality.

\section{Preliminaries}\label{sec: Preliminaries}

In this section, we  introduce the compatibility-hypergraph (CH) approach to contextuality \cite{amaral2018graph}, which is the framework we will work with. More precise definitions can be found in appendix \ref{appendix: compatibility-hypergraph}.

The starting point of the CH approach is the idea of \textbf{measurement scenario} (definition~\ref{def: Scenario}). It consists in a finite set $\mathcal{A}$ representing measurements that can be performed on some physical system, together with a collection $\mathcal{C}$ of maximal contexts. A context is a set $C  \subset \mathcal{A}$ representing compatible measurements, i.e., measurements that can be  performed jointly, and it is required that the collection $\mathcal{C}$  satisfies the following conditions: (1) Any $C \in \mathcal{C}$ is maximal, i.e., if  $C',C''\in \mathcal{C}$ satisfy $C'' \subset C'$ then $C' = C''$, and (2) each measurement belongs to at least one context, that is, $\cup \mathcal{C} = \mathcal{A}$. Denoting by $O$ the set of possible outcomes for all measurements, we say that the triple of sets $\mathcal{S} \equiv (\mathcal{A},\mathcal{C},O)$ is a measurement scenario \cite{amaral2018graph, Abramsky_2011}. 

If the system is prepared in a certain way and all measurements of a context $C$ are performed simultaneously, a joint outcome, which we represent as a function $C \ri O$, will be obtained. If we prepare the system in exactly the same way  -- by ``exactly the same'' we mean that the agent cannot distinguish both preparation procedures -- and run the experiment a second time, we will not necessarily obtain the same joint outcome. It means that a preparation procedure does not determine an outcome for each measurement of $C$ but it actually defines a probability distribution $p(\cdot| C)$ on the set $O^{C}$ of all functions $C \ri O$; this probability distribution is experimentally accessible by means of successive runs of the experiment. This is true for any context $C \in \mathcal{C}$, therefore a preparation procedure defines a mapping $C \xrightarrow{p} p(\cdot | C)$ associating  to each context $C$ a probability distribution $p(\cdot| C)$ on $O^{C}$. This mapping $p$ is said to be a \textbf{behavior} (definition \ref{def: behavior}) or empirical model \cite{Abramsky_2011} in the measurement scenario $\mathcal{S}$.

A behavior $p$ in a scenario $\mathcal{S}$ is said to be \textbf{noncontextual} (definition \ref{Def: Noncontextual}) if a probability distribution $\overline{p}$ on $O^{\mathcal{A}}$ exists such that, for any context $C$, its marginalization on $O^{C}$ coincides with $p(\cdot| C)$; otherwise $p$ is said to be contextual \cite{amaral2018graph}. The set of all noncontextual behaviors in a scenario $\mathcal{S}$ defines a polytope, so it can be characterized by finitely many linear inequalities; a \textbf{noncontextuality inequality} is a (nontrivial) linear inequality characterizing the polytope of noncontextual behaviors on $\mathcal{S}$; it is satisfied by all noncontextual behaviors but violated by some contextual ones. We suggest Ref.~\cite{amaral2018graph} for an explicit definition. The important point is that a noncontextuality inequality is violated by a behavior $p$ only if $p$ is contextual, whereas a contextual behavior necessarily violates some noncontextuality inequality of the scenario. Bell inequalities are particular cases of noncontextuality inequalities \cite{amaral2018graph}.

A \textbf{n-cycle} is a scenario $\mathcal{S}_{n}$ containing $n$ dichotomic measurements $A_{0}$, $\dots$, $A_{n-1}$ and $n$ maximal contexts $C_{i} \equiv \{A_{i},A_{i+1}\}$, $i=0,\dots, n-1$ (if $i=n-1$, then $i+1 \equiv 0$) \cite{AraujoCycle, amaral2018graph}. Given a behavior $p$ on $\mathcal{S}_{n}$, we will sometimes denote its component $C_{i}$ by  $p(\cdot | A_{i},A_{i+1})$ rather than $p(\cdot | C_{i})$. Note that only four joint outcomes are possible in context $C_{i}$, namely $(\bot,\bot), (\bot,\top),(\top,\bot)$ and $(\top,\top)$. We are interested in the following important example of (maximally) contextual behavior in the $5$-cycle \cite{amaral2018graph}:
\begin{defn}[generalized coin toss]\label{def: n-faced coin toss} Let $\mathcal{S}_{5}$ be the 5-cycle. A ``generalized coin toss'' on $\mathcal{S}_{5}$ is the behavior $p$  such that, for any context $C_{i} \equiv \{A_{i},A_{i+1}\}$, $p(\cdot | C_{i}) \equiv p(\cdot | A_{i},A_{i+1})$ is given by
\begin{align}
    p(\bot,\bot|A_{i},A_{i+1}) &= 0 =  p(\top,\top|A_{i},A_{i+1}),
    \\
    p(\bot,\top|A_{i},A_{i+1}) &= \frac{1}{2} =  p(\top,\bot|A_{i},A_{i+1}).
\end{align}
\end{defn}
Straightforward calculations show that this behavior is nondisturbing (definition \ref{Def: Nondisturbance}). As a consequence, $p$ associates, for any measurement $A_{i}$, a unique (i.e., context-independent) probability distribution $p(\cdot| A_{i})$ on $\{\bot,\top\}$ --- to define $p(\cdot| A_{i})$, just take any component of $p$ containing $A_{i}$ and marginalize it. All measurements $A_{i}$, $i=0,\dots,4$, have the same distribution, namely
\begin{align}\label{eq: marginal}
    p(\bot|A_{i}) = \frac{1}{2} = p(\top|A_{i}).
\end{align}
It is well known that this behavior has no quantum realization (definition \ref{def: Quantum behavior}) \cite{amaral2018graph}; in particular, it is contextual (definition \ref{Def: Noncontextual}). Finally, it violates the \textbf{KCBS inequality}, a famous noncontextuality inequality for the $5$-cycle \cite{amaral2018graph, CSW}, named after Alexander A. Klyachko, M. Ali Can, Sinem Binicioğlu and  Alexander S. Shumovsky \cite{KCBSeminal}, which asserts (by means of an equivalence \cite{CSW, amaral2018graph}) that a behavior $p$ in the $5$-cycle is noncontextual only if it satisfies
\begin{align}
    \sum_{i=0}^{4} p(\bot,\top\vert A_{i},A_{i+1}) \leq 2.
\end{align}
For the generalized coin toss we have $\sum_{i=0}^{4} p(\bot,\top\vert A_{i},A_{i+1}) = \frac{5}{2} >2$, hence the KCBS inequality is (maximally) violated by it \cite{amaral2018graph, CSW}. 

\section{Classicality and classical realism}\label{sec: classicality}

An important feature of the theory-independent definition of noncontextuality presented in section \ref{sec: Preliminaries} is that a behavior $p$ in $\mathcal{S} \equiv (\mathcal{A},\mathcal{C},O)$ is noncontextual (definition \ref{Def: Noncontextual}) if and only if a classical realization (definition \ref{def: Classical behavior}) exists for $p$. It means that $p$ is noncontextual if and only if it is possible to represent all measurements of $\mathcal{S}$ as random variables in a probability space $(\Lambda,\Sigma,\mu)$ in such a way that, for any context $C$, $p( \ \cdot \ \vert C)$ is the joint probability distribution of all random variables in $C$. Clearly, a classical realization is closely related to the  description of a physical system in classical mechanics: the measurable space $(\Lambda,\Sigma)$ is like a state space, measurements are functions on this ``state space'', just as in the classical case, and the preparation procedure determining $p$ is a probability measure $\mu$ on $(\Lambda,\Sigma)$, in analogy to classical mechanics, where states are represented by probability measures on the state space. If by ``non-classicality'' one means the obstruction to the existence of classical realizations, then contextuality is, by definition, a signature of non-classicality. However, inspired by Ref. \cite{WhatThing}, we say that a classical realization is a \textbf{classical \textit{realist} description}, and we distinguish classical realism from classicality. As we mentioned in the introduction, we believe that a signature of \textbf{non-classicality} must tell us when ``nature fails to respect classical physics'' \cite{NatureMazurek}, and, as we will argue throughout the paper, the obstruction to classical realist descriptions does not satisfy, as we see it, this necessary condition for non-classicality. 

For the sake of clarity, let's define ``classical realism'' in a more precise way.

\begin{defn}[Classical realism]\label{def: classical realism} We say that a behavior $p$ 
 in a scenario $\mathcal{S}$ admits a classical realist description if it is compatible with the three following views. 
\begin{itemize}
    \item[(a)] \textbf{Factualism}: There is a set $\Lambda$ of possible ``hidden states'' for the system that $p$ describes, and, at any given time, the \textit{state of affairs} of this system is determined by a single hidden state.
\item[(b)] \textbf{Ontic view on measurements:}  Any measurement $A$ of the scenario is associated with a property (also denoted by $A$) that is fully specified by the hidden state of the system, and a joint measurement of a context $C$ simply reveals the values these properties possess at that moment.
\item[(c)] \textbf{Epistemic view on preparation procedures:} The preparation procedure $P$ that gives rise to the behavior $p$ encodes, and is uniquely determined by, a certain degree of knowledge about the hidden state of the system. 
\end{itemize}
\end{defn}

A \textit{state of affairs} is a ``truth-maker'', i.e., an assignment of truth values for all propositions about the properties of the system \cite{WhatThing}. Item $(a)$ of definition \ref{def: classical realism} asserts that the hidden state of the classical realization, being a state of affairs, determines ``the way things are at a particular moment of time'', as Isham and Döring put it \cite{WhatThing}. Thus, according to this definition, the hidden state determines all the \textit{facts} about the (properties of the) system. This is why we use the term ``factualism'', which bears a similar meaning in philosophy \cite{sep-states-of-affairs, armstrong1993world}. The terms ``ontic'' and ``epistemic'' in items $(b)$ and $(c)$ are motivated by Spekkens' approach to ontological models \cite{Spekkens_2007, SpekkensSeminal, LeiferIs}, but all we mean by them is what was stated in definition \ref{def: classical realism}. 

Up to mathematical subtleties, $p$ admits a classical realist description if and only if $p$ is noncontextual. In fact, saying that a property $A$ is fully specified by the hidden state of the system means that each hidden state $\lambda \in \Lambda$ ascribes a definite value to $A$, which in turn means that $A$ defines a function $f_{A}$ on $\Lambda$. On the other hand, a mathematically rigorous way of putting item $(c)$ is by saying that a preparation procedure $P$ defines a probability measure $\mu_{P}$ on $\Lambda$ (endowed with an appropriate $\sigma$-algebra $\Sigma$) in such a way that, for any measurable set  $U \subset \Lambda$, $\mu_{P}(U)$ is the probability that the state of the system lies in $U$ \cite{Kochen1975}. Finally, items $(b)$ and $(c)$ together imply that $p( \ \cdot \ \vert C)$ is the joint distribution of $\{f_{A}: A \in C\}$, where, clearly, we are assuming that $f_{A}$ is a measurable function on $\Lambda$ for every measurement $A$. 

Kochen-Specker theorem, by which we mean the non-existence of valuation functions on the set of bounded selfadjoint operators of a Hilbert space \cite{Kochen1975, doring2005kochen}, brings out a  tension between factualism and the ontic view on projective measurements in quantum theory, insofar it shows that we cannot consistently interpret all projective measurements (or, equivalently, all selfadjoint operators) as representing properties simultaneously possessed by the system \cite{Kochen1975, doring2005kochen, WhatThing}.  On the other hand, some of the so-called state-dependent proofs of quantum contextuality \cite{budroni2021quantum} (including some violations of Bell inequalities) rule out classical realist descriptions for quantum behaviors in scenarios where valuation functions can in principle be defined, obstructing classical realism without explicitly appealing to a contradiction between factualism and the ontic view on measurements. Hence, all three aspects of classical realism presented in definition \ref{def: classical realism} are in check in quantum theory \cite{budroni2021quantum}.  

\section{Inside the  Device}\label{section: Inside the  box}
The toy mechanism we propose is an input/output device operating along the lines set out in Refs.~\cite{barbosa2021closing, AmaralWirings}, and it gives rise to the contextual behavior we named ``generalized coin toss''. The idea that an experimental setup can be seen as an input/output device or a ``black box'' has been dominating the debate on quantum foundations, and contextuality is no exception \cite{Abramsky_2011, AmaralWirings, barbosa2021closing}. As Sandu Popescu puts it (when discussing non-locality), ``one can imagine that inside the box there is an automated laboratory, containing particles, measuring devices, and so on. The laboratory is prearranged to perform some specific experiments; the input $x$ simply indicates which experiment is to be performed[...] In this framework, the entire physics is encapsulated in $P(a|x)$, the probability that output $a$ occurs given that measurement $x$ was made'' \cite{NaturePopescu}. Note that one of the core principles of classical realism (definition \ref{def: classical realism}) is explicitly left aside here, namely the ontic view on measurements. 

\begin{figure}[h]
\centering
\begin{subfigure}{.45\textwidth}
  \centering
  \includegraphics[width=1\linewidth]{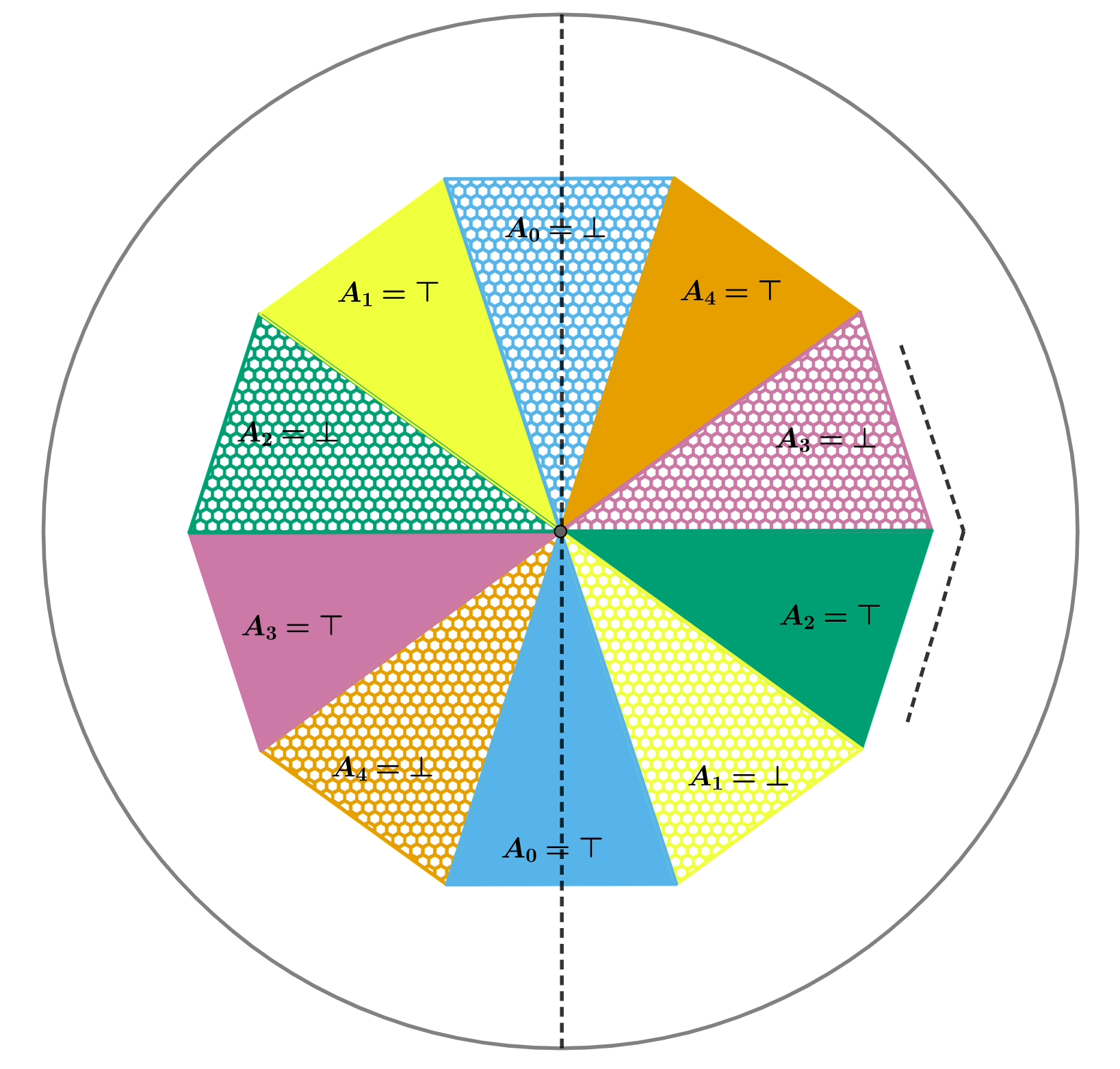}
  \caption{Object and context $C_{2} \equiv \{A_{2},A_{3}\}$}
  \label{fig: object}
\end{subfigure}%
\begin{subfigure}{.45\textwidth}
  \centering
  \includegraphics[width=1\linewidth]{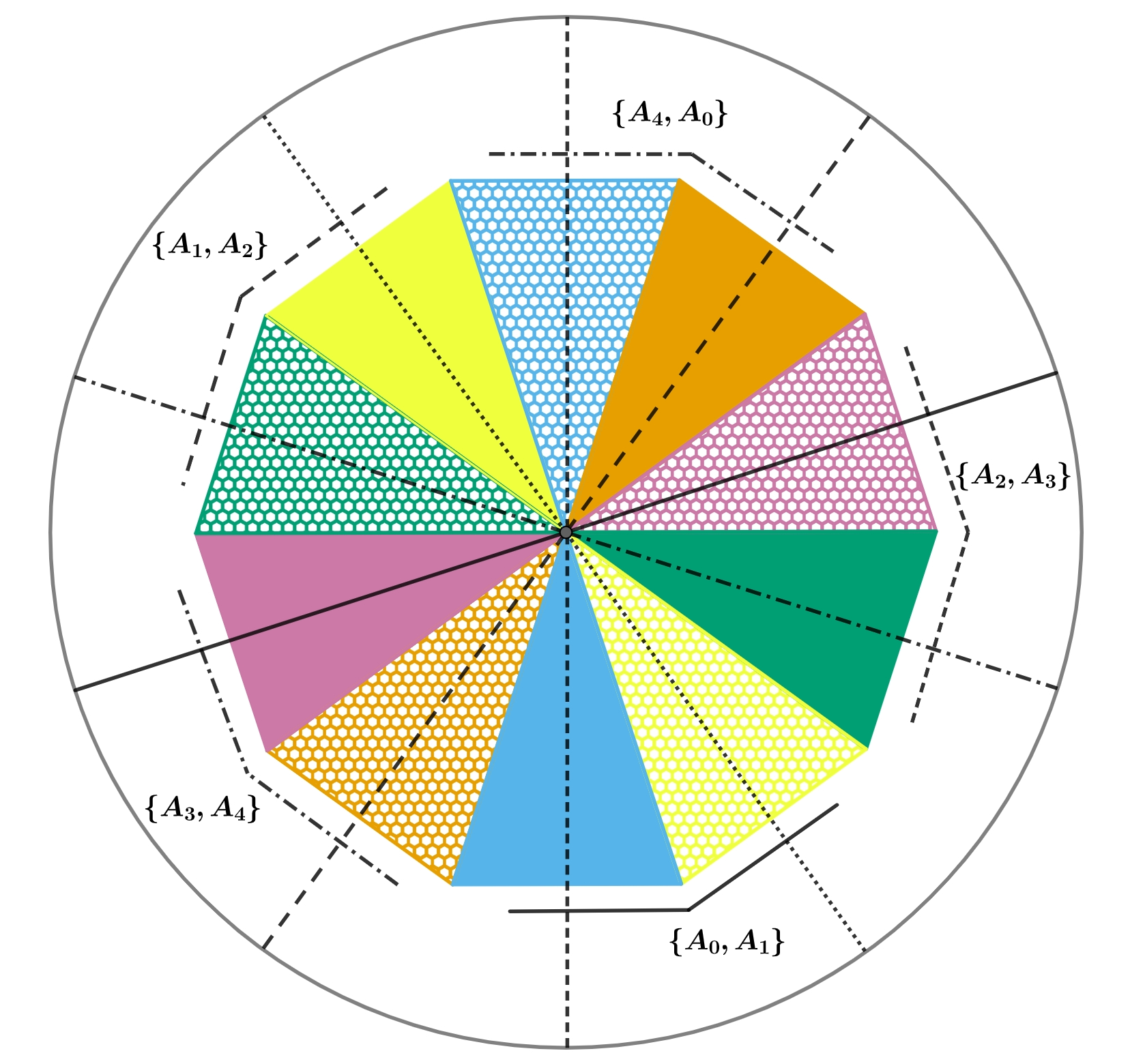}
  \caption{All axes and detectors}
  \label{fig: mechanism}
\end{subfigure}
\caption{Schematic representation of the object (figure \ref{fig: object}) and the toy mechanism (figure \ref{fig: mechanism})}
\label{fig:test}
\end{figure}

As we know, a coin toss is a measurement procedure that, by means of a convention, returns one out of two previously fixed characteristics (opposite sides) of a coin. Given that, for all practical purposes, a coin has only one pair of opposite sides,  a coin toss can be thought of as a measurement procedure where only one dichotomic measurement is performed, and it is commonplace in operational approaches to physics to discuss measurement procedures that are operationally indistinguishable from a coin toss --- see, for instance, Refs.~\cite{NatureMazurek, SpekkensSeminal}. In a similar fashion, by conceiving an object having $n$ pairs of opposite sides, we can use this object to perform $n$ distinct epistemic measurements, one for each pair of opposite sides. We represent this object as a regular polygon of $2n$ sides, and figure \ref{fig: object} depicts it for $n=5$.

A coin toss can be idealized as a rotation around a fixed axis passing through the center of the coin, and so will our measurements; the only difference is that, in the case of a ``generalized coin toss'', we must take contexts into account. In our device, what determines which measurements can be performed simultaneously is a collection of axes around which the classical object can rotate, together with detectors that are capable of identifying colors and distinguishing between light and dark colors. We have five axes and five detectors, each detector being associated with a single axis, and a measurement in our device consists in a ``coin toss'', i.e., a rotation around one of the axis, followed by a detection in the detector associated with it. The mechanism is depicted in Figure \ref{fig: mechanism}, where axes are represented as straight lines passing through the center of the object, and detectors are represented by the segments surrounding the object. Each pair ``axis + detector'' is identified by a type of dashed line. Each measurement (i.e., each pair of opposite sides) is characterized by a specific color and a specific label $A_{i}, i=0,\dots,4$. Each color has two shades, a dark one for the outcome $\top$ and a light one for the outcome $\bot$; to make the distinction between dark and light clearer in the figures, we fill the light side of all measurements using a hexagonal pattern (see figure~\ref{fig: object}).  Note that only pairs of consecutive measurements (i.e., pairs $(A_{i},A_{i+1})$) can be measured jointly.

\begin{figure}[!htb]
    \centering
    \includegraphics[width=9cm]{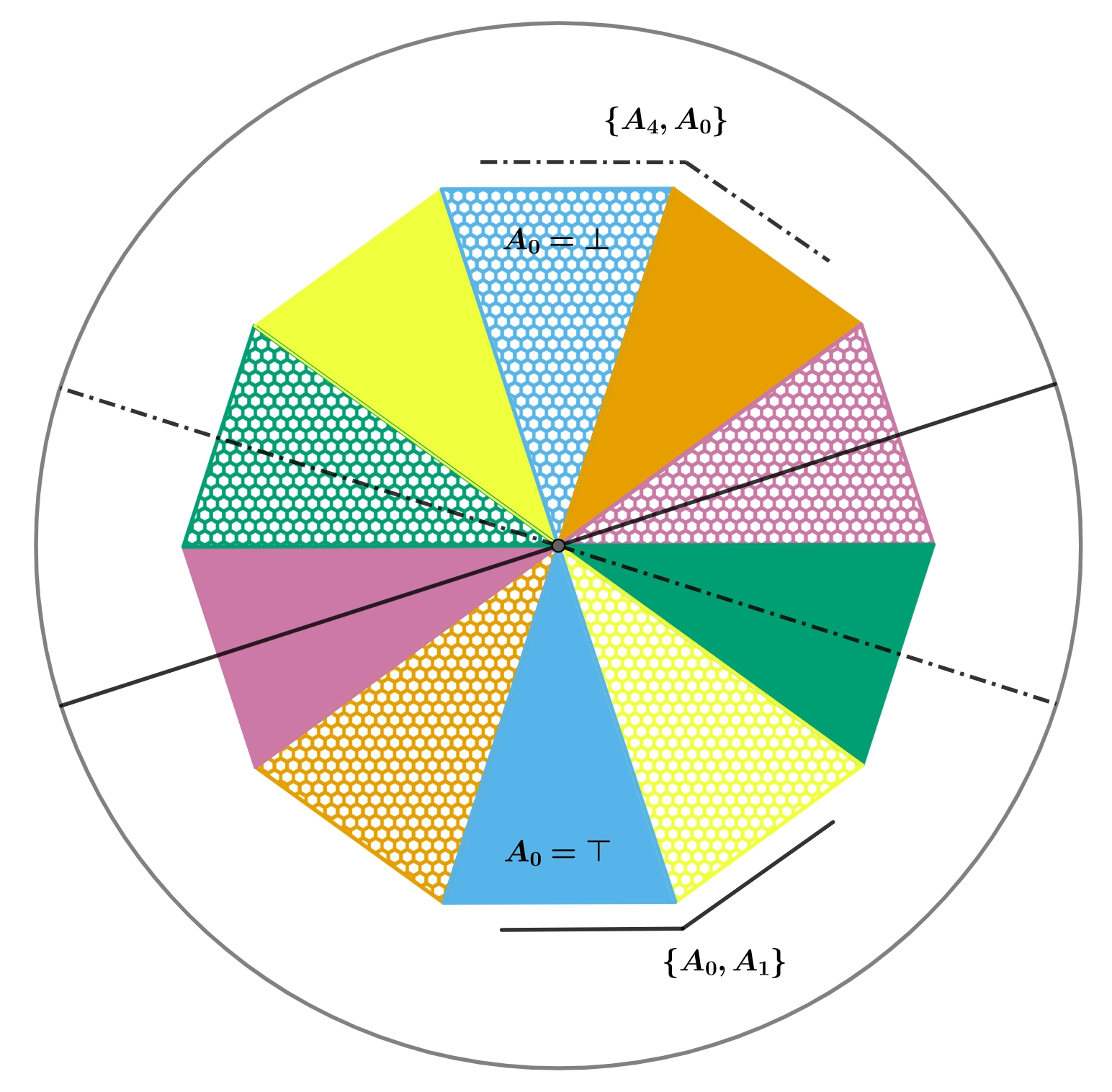}
    \caption{Schematic representation of the pair of contexts containing measurement $A_{0}$ (blue)
    }
    \label{fig: C0C4}
\end{figure}

In our thought experiment, this toy mechanism lies inside a black box, which means that the agent only has access to the data this device generates. Following the description of black boxes for contextuality analysis described in Ref.~\cite{AmaralWirings}, we can imagine that the black box has buttons, one for each measurement, and that pressing a button is equivalent to performing a measurement.  If a single button, say $A_{0}$, is pressed, one of the two detectors that are capable of detecting this color at that moment  (see figure \ref{fig: C0C4}) will be randomly selected by the mechanism --- alternatively, we could let the agent choose the detector ---, and a ``coin toss'' around the axis associated with this detector will be performed, which means that an uncontrolled kick will rotate the object around it. At the end of the movement, the detector will return $(\top|A_{0})$ in case it detects ``dark blue'', and $(\bot|A_{0})$ in case it detects ``light blue''; the probability of both these events is $\frac{1}{2}$ because the measurement procedure is an idealized coin toss. The same statistics will be obtained for any single measurement. Similarly, if we press a pair of consecutive buttons simultaneously, i.e., if we perform a joint measurement within a context, a ``coin toss'' around the axis associated with the detector capable of detecting these colors will be performed. If we measure $A_{2}$ and $A_{3}$ jointly, for instance, the detector will return $(\top, \bot|A_{2},A_{3})$ in case it detects ``dark green'' and ``light purple'', and $(\bot, \top|A_{2},A_{3})$ in case it detects ``light green'' and ``dark purple''; note that, thanks to a mere geometrical property of the object, these are the only possible joint outcomes for this joint measurement (see figure \ref{fig: object}), and the analogous result holds for any context. Whenever we perform a joint measurement $A_{i},A_{i+1}$ in this device, the probability of obtaining $(\top,\bot|A_{i},A_{i+1})$ is half, just as the probability of obtaining $(\bot,\top|A_{i},A_{i+1})$; the probability of obtaining $(\top,\top|A_{i},A_{i+1})$ or $(\bot,\bot|A_{i},A_{i+1})$ is zero. Thus, if all we are able to do is to perform this context-dependent collection of joint measurements upon this simple classical system, the data we will obtain will be, in the limit of many trials, the important example of contextual behavior we named ``generalized coin toss'' (definition \ref{def: n-faced coin toss}). Therefore, the data generated by this black box violates the KCBS inequality.

\section{Discussion}

In section \ref{sec: classicality} we argued that a behavior is noncontextual if and only if a classical realist description exists for it, and we defined classical realism in terms of three individually necessary and conjointly sufficient conditions, namely $(a)$ factualism, $(b)$ the ontic view on measurements and $(c)$ the epistemic view on preparation procedures --- see definition \ref{def: classical realism}.  In the toy mechanism we proposed, hidden states for the system inside the box do exist, since it is a  rigid body that obeys the laws of classical mechanics, and our preparation procedure does encode, and is uniquely determined by, a certain degree of knowledge about these states, as in the standard coin toss. It means that both factualism and the epistemic view  on preparation procedures are compatible with our toy mechanism. On the other hand, our measurements do not reveal preexisting values of properties because they are not measuring properties of the system at all --- just as no property of a coin is measured when a coin is tossed ---, thus item $(b)$ of definition \ref{def: classical realism} is not satisfied, and this is why contextual data emerge from our mechanism. Although it is hard to imagine a classical system (i.e., a system obeying the laws of classical physics) that violates items $(a)$ or $(c)$ of definition \ref{def: classical realism}, we see no connection between item $(b)$ and the necessary condition for non-classicality we mentioned in section \ref{sec: classicality}, namely that ``nature fails to respect classical physics'' \cite{NatureMazurek}, since the ontic view on measurements depends on the experimental setup as a whole, and not only on the system itself. In short, the ontic view on measurements (or, more specifically, the link between measurements and properties) is about realism, not about physical phenomena. Therefore, it is not a surprise that, by being exceedingly permissive with the notion of measurement procedure, we can generate data that have no classical \textit{realist} description  using a device whose  underlying structure in no way ``fails to respect classical physics''. Contextuality defies ``classical understanding'' \cite{NatureMazurek} especially when measurements are thought of as revealing values of properties, i.e., when some sort of realism is at stake: under this assumption, we have to appeal to counter-intuitive and non-classical explanations for contextuality such as ``properties that have no preexisting values'', ``properties whose values depend on other jointly measurable properties'', ``contextual hidden states'', and all other well known intricate concepts we have to deal with when discussing contextuality in quantum theory \cite{WhatThing, halvorson2019realist}, where the connection between measurements and properties plays a central role \cite{Kochen1975, budroni2021quantum, WhatThing}. Once we reject realism \textit{a priori} by deliberately disconnecting measurements from properties, the counterintuitiveness of contextuality, and consequently the necessity of non-classical explanations for it, turn out to be less evident. As we mentioned, leaving the ontic view on measurements aside is commonplace in theory-independent approaches to physics, so our thought experiment provides a very simple illustration of how this choice can impact the relationship between contextuality and non-classicality. In particular, it illustrates how important the definition of measurement is for the phenomenological significance of contextuality, as already pointed out by other authors \cite{ZhangExperimental}.  

As we discussed in section \ref{sec: classicality}, the reason for  representing measurements as random variables in hidden variable models is the assumption that they are associated with properties whose values are completely determined by the hidden state of the system: the random variable $f_{A}$ representing the measurement/property $A$ assigns each hidden state $\lambda$ to the value possessed by  $A$ when the state of the system is $\lambda$ \cite{Kochen1975, WhatThing, doring2005kochen}. If we assume factualism but reject the ontic view on measurements, there is no reason to require measurements to be represented as functions in hidden variable models, since only properties need to have definite values at all times. The outcome of a coin toss, for instance, is determined only when the coin is flipped, and it doesn't make sense to ask whether the coin I have in my pocket right now is heads or tails. Points of a measurable space do not assign values to Markov Kernels, so replacing random variables with Markov Kernels in definition \ref{def: Classical behavior} turns it into a definition that, in principle, also deserves to be labeled ``classical''. This is the well-known assumption of outcome indeterminism \cite{SpekkensSeminal}. 

In quantum theory, there is no distinction between joint and sequential measurements within contexts \cite{tezzin2023stepbystep}, and joint measurements in experimental tests of quantum contextuality can be performed sequentially \cite{budroni2021quantum}. On the other hand, frameworks for theory-independent contextuality are usually not designed to accommodate sequential measurements \cite{Abramsky_2011, amaral2018graph, budroni2021quantum}, and joint measurements are assumed to be simultaneous in them. Our toy mechanism is an example of a black box where joint and sequential measurements within contexts are not equivalent: in a sequential measurement of $A_{i},A_{i+1}$, the sequential outcomes $(\top,\top)$ and $(\bot, \bot)$ can be obtained, whereas in a joint measurement of $A_{i},A_{i+1}$ it is impossible. Furthermore, the behavior $p$ that we generate by performing sequential measurements within contexts with our black box satisfies $p(\top,\top\vert A_{i},A_{i+1})= p(\top,\bot\vert A_{i},A_{i+1}) = p(\bot,\top\vert A_{i},A_{i+1}) = p(\bot,\bot\vert A_{i},A_{i+1}) = 1/4$ for any $i=0,\dots,4$, and it is easy to see that it is noncontextual. Hence, the contextuality status of the data generated by our back box depends on whether measurements are being performed simultaneously or sequentially, which is only possible because our measurements are not ideal \cite{tezzin2023stepbystep, cabello2019simple}. As we see it, it indicates that the equivalence between joint and sequential measurements within contexts, which is assured when we consider sequential measurements of ideal observables \cite{tezzin2023stepbystep}, may be relevant for contextuality analysis in other fields, as it is in quantum theory. 

Most known examples of systems that are classical in some sense but generate contextual data lie outside the realm of physics, as in psychology or linguistics \cite{cervantes2018snow, Atmanspacher2019, wang2021analysing}. Despite being important for a variety of reasons, these examples usually do not aim to question the phenomenological status of contextuality in physics, due both  to their intentionally ``non-physical'' nature and to the fact that the data they give rise to are usually disturbing (definition \ref{Def: Nondisturbance}), requiring a ``generalized'' definition of contextuality in order to be analyzed \cite{CbD2.0, Matt2019}. Our toy mechanism, on the other hand, is intentionally designed to fit into the black box description of experiments that is commonly found in the physics literature \cite{Abramsky_2011, AmaralWirings, barbosa2021closing}, and the data our device generates is nondisturbing. In particular, the view put forward here can be relevant for the debate about contextuality in physical systems which are not strictly quantum, like classical light \cite{li2017experimental, li2019state}, especially with regard to the distinction between classical realism and classicality. Furthermore, the data generated by our black box is, as we mentioned, an important example of contextual behavior that maximally violates the KCBS inequality and for which no quantum assignment (definition \ref{def: Quantum behavior}) can be conceived  \cite{amaral2018graph}, thus having this specific example as a target is particularly relevant. Finally, one of the main reasons why the KCBS inequality is important is that it cannot be turned into a Bell inequality \cite{cabello2021converting}, so the fact that our mechanism violates it distinguishes it from toy models for experimental data in Bell scenarios \cite{Hofer2021Causal}.

\begin{figure}[htbt]
    \centering
    \includegraphics[width=9cm]{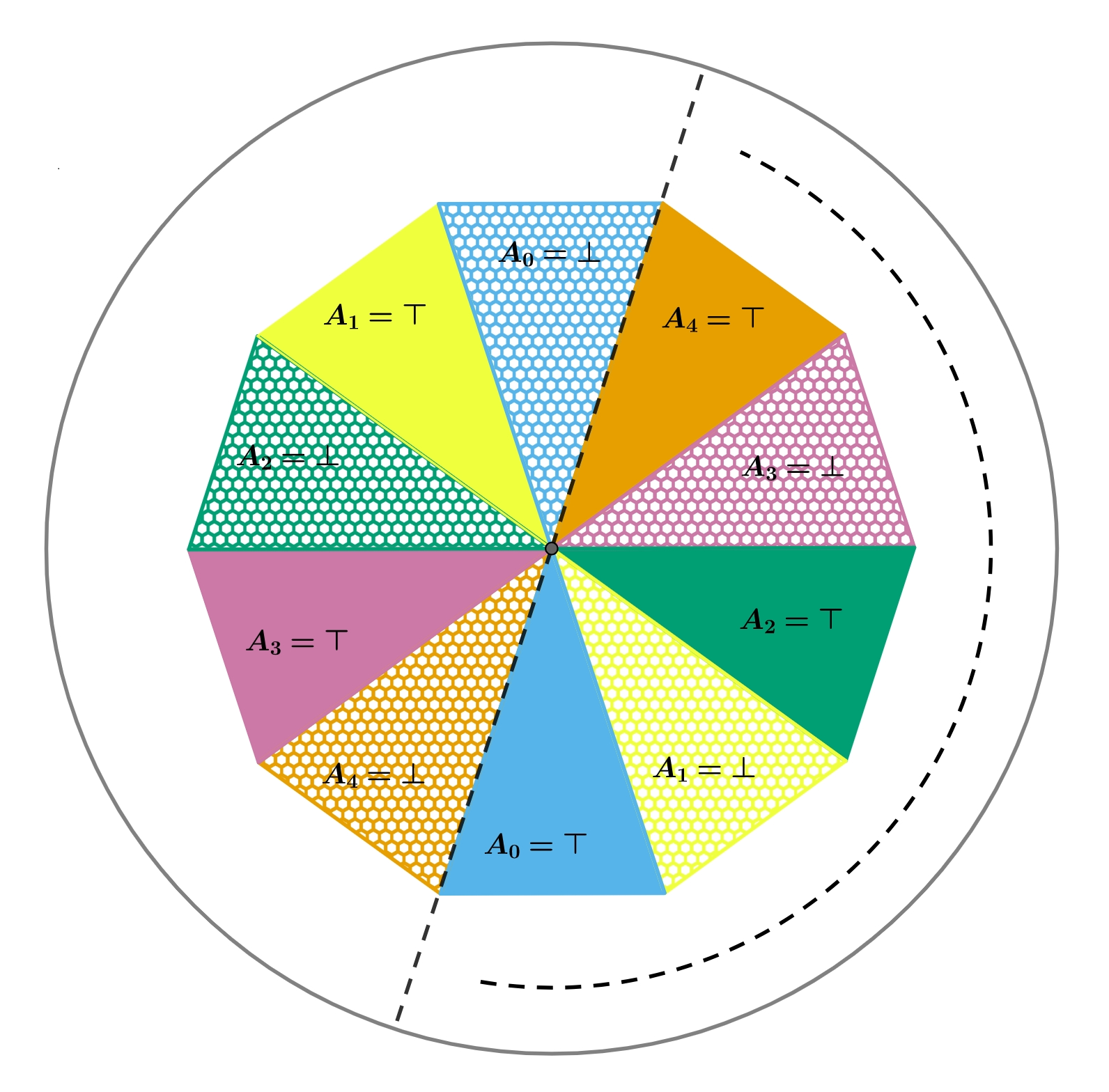}
    \caption{Schematic representation of a rearrangement of the device without context-dependent outcomes
    }
    \label{fig: finalContext}
\end{figure}

To conclude, we would like to discuss the connection between contextuality and context-dependence of outcomes in our thought experiment. Let $A_{i}$ be any measurement and $C_{i}, C_{i-1}$ be the pair of contexts containing it. It makes sense to say that, according to this device, given a final configuration of the object, contexts $C_{i}$ and $C_{i-1}$ return distinct outcomes for $A_{i}$, because, as illustrated for in figure \ref{fig: C0C4}, the detectors associated to these contexts lies in opposite sides of the object. For this reason, we can say that, in this device, the outcomes of measurements depend not only on the measurement itself but also on the context we choose. It is worth mentioning that we can make such a statement just because we have access to the underlying structure of the device; if we treat it as a black box, then we can only relate different contexts using counterfactual reasoning like ``if, instead of using context $C_{i}$, we had used context $C_{i+1}$, would we have obtained the same outcome?'', and the fact that the data generated by the device is probabilistic leaves no room for affirmative answers for such questions. In any case, one can ask whether it is possible to rearrange the device in such a way that outcomes cease to be context-dependent, while, at the same time, the data generated by it remain contextual. As one should expect, it is not possible. In fact, in order to get rid of context-dependence we need to overlap all detectors containing $A_{i}$ for all $i=0,\dots,4$, as illustrated in figure \ref{fig: finalContext} (all other possible choices leads to the same conclusion). Note that, for the sake of simplicity, we have chosen the same axis for all contexts, and we have represented the overlapping detectors as a single one, so that, when we select a measurement (or a pair of consecutive measurements), we simply inform the box what is the color we are interested in. The fact is that, with this kind of configuration, the data generated by the device is not the ``generalized coin toss'' anymore. In the configuration we have chosen, for instance, we obtain $p(\top,\bot|A_{i},A_{i+1}) = \frac{1}{2} = p(\bot,\top| A_{i},A_{i+1})$ for $i=0,\dots 3$ but $p(\top,\top|A_{4},A_{0}) = \frac{1}{2} = p(\bot,\bot| A_{4},A_{0})$. This behavior is noncontextual, having
\begin{align}
    p(\top,\bot,\top,\bot,\top|A_{0},A_{1},A_{2},A_{3},A_{4}) &= \frac{1}{2}
    \\
    &= p(\bot,\top,\bot,\top,\bot|A_{0},A_{1},A_{2},A_{3},A_{4})
\end{align}
as a global section. Finally, it is easy to see that we can only generate the ``generalized coin toss'' if, for at least one measurement $A_{i}$,  outcomes are context-dependent, i.e., the detectors associated with $C_{i}$ and $C_{i-1}$ lies in opposite sides of the object. 

\section*{Acknowledgments} \label{sec:acknowledgements}
I would like to thank Bárbara Amaral, Rafael Wagner, Giulio Halisson, and Leonardo Santos for insightful discussions and helpful comments. I also would like to thank anonymous reviewers for their comments and criticisms, which significantly improved this paper. This preprint has not undergone peer review or any post-submission improvements or corrections. The Version of Record of this
article is published in Foundations of Science, and is available online at \url{https://doi.org/10.1007/s10699-023-09928-7}.

\section*{Funding}
This work was funded by  National Council for Scientific and Technological Development (CNPq).

\section*{Statements and Declarations}
The author has no competing interests to declare that are relevant to the content of this article. 

\section*{Data Availability}
Data sharing not applicable to this article as no datasets were generated or analyzed during the current study.

\appendix
\section{Appendix}\label{appendix: compatibility-hypergraph}
In this appendix we provide a brief overview of the compatibility-hypergraph approach to contextuality.
Our main reference is Ref.~\cite{amaral2018graph}.

\begin{defn}[Scenario]\label{def: Scenario} A scenario is a triple $\mathcal{S} \equiv (\mathcal{A},\mathcal{C},O)$ where $\mathcal{A}$, $O$ are finite sets, whose elements represent measurements and outcomes respectively, and $\mathcal{C}$ is a collection of subsets of $\mathcal{A}$ (each one representing a maximal set of compatible measurements) satisfying the following conditions.
\begin{itemize}
\item[(a)] $\mathcal{A} = \cup \mathcal{C}$ \label{property a}
\item[(b)] For $C,C' \in \mathcal{C}$, $C' \subset C$ implies $C' = C$
\end{itemize}
\end{defn}
The approach is named "hypergraph-approach" because a scenario $(\mathcal{A},\mathcal{C},O)$ can be associated with a hypergraph whose vertices are the elements of $\mathcal{A}$ and whose hyperedges are the elements of $\mathcal{C}$ \cite{amaral2018graph}. 

The result of a joint measurement over a context $C$ can be represented as a function $C \ri O$. Therefore, the set $O^{C}$ of all functions $C \ri O$ can be understood as the set of all possible outcomes of a joint measurement on $C$. Behaviors enable us to encode  experimental data obtained from joint measurements on a measurement scenario \cite{amaral2018graph}.

\begin{defn}[behavior]\label{def: behavior}
Let $\mathcal{S}$ be a scenario. A behavior on $\mathcal{S}$ is a function $p$ which associates to each context $C$ a probability distribution $p(\cdot | C)$ on $O^{C}$, that is to say, for each context $C$, $p(\cdot | C)$ is a function $O^{C} \rightarrow [0,1]$ satisfying $\sum_{s \in O^{C}} p(u | C) = 1$. 
\end{defn}

A behavior whose components coincide when restricted to intersections of contexts is said to be non-disturbing \cite{amaral2018graph}: 
\begin{defn}[Non-disturbance]\label{Def: Nondisturbance}
A behavior $p$ in a scenario $\mathcal{S}$ is said to be non-disturbing if, for any pair of intersecting contexts $C,D$, equality 
\begin{align}p(\cdot | C \cap D, C) = p(\cdot | C \cap D, D)
\end{align}
holds true, where, for any $E \in \mathcal{C}$ and $E' \subset E$, $p(\cdot | E',E)$ denotes the marginal of $p(\cdot | E)$ on $O^{E'}$, namely
\begin{align}
  \forall t \in O^{E'}: \ \ \ p(t| E', E) \doteq\sum_{\begin{subarray}{l} s \in O^{E} \\ s|_{E'} = t \end{subarray}} p(s|E).
\end{align}
\end{defn}

Classical and quantum behaviors arise when ideal measurements are performed upon classical and quantum systems respectively. They are defined as follows \cite{amaral2018graph}.
\begin{defn}[Classical realization]\label{def: Classical behavior} Let $p$ be a behavior in a scenario $\mathcal{S} \equiv (\mathcal{A},\mathcal{C},O)$. A probability space $\boldsymbol{\Lambda} \equiv (\Lambda,\Sigma,\mu)$ is said to be a classical realization for $p$ if we can associate each measurement $A$ of $\mathcal{S}$ to a random variable $f_{A}: \Lambda \ri O$ in $\boldsymbol{\Lambda}$ is such a way that, for any context $C$, $p(\cdot |C)$ is the joint distribution of the set  $\{f_{A}: A \in C\}$,which means that, for any $s \in O^{C}$,
\begin{align}
    p(s|C) = \mu\left( \bigcap_{A \in C} f_{A}^{-1}(\{s_{A}\}) \right),
\end{align}
where $s_{A} \equiv s(A)$. 
\end{defn}
The definition of quantum realization goes as follows \cite{amaral2018graph}.

\begin{defn}[Quantum realization]\label{def: Quantum behavior} Let $p$ be a behavior in a scenario $\mathcal{S} \equiv (\mathcal{A},\mathcal{C},O)$. A pair $(H,\rho)$, where $H$ is a finite-dimensional Hilbert space and $\rho$ is a density operator on $H$, endowed with a mapping $\mathcal{A} \ni A \xmapsto{\Theta} \Theta(A) \equiv T_{A} \in \mathcal{B}(H)_{\text{sa}}$, is said to be a quantum realization for $p$ if the following conditions are satisfied.
\begin{itemize}
    \item[(a)] $O$ is, up to isomorphism, the set of eigenvalues of $\Theta(\mathcal{A})$, that is to say,
    \begin{align}
        O \cong \bigcup_{A \in \mathcal{A}} \sigma(T_{A}).
    \end{align}
    \item[(b)] Each context is embedded by $\Theta$ into a commutative algebra, i.e., if $A,B \in C$ for some context $C$, then $T_{A}$ and $T_{B}$ commute.
    \item[(c)] For each context $C$, $p(\cdot| C)$ is reproduced by the Born rule, i.e., for any $s \in O^{C}$,
    \begin{align} 
    p(s|C) = \text{Tr}\left(\rho \prod_{A \in C} P^{(A)}_{s_{A}}\right),
    \end{align}
    where $P^{(A)}_{s_{A}}$ denotes the projection $\chi_{\{s_{A}\}}(A)$ (if $s_{A} \in \sigma(T_{A})$, this is the projection associated with the subspace of $H$ spanned by the eigenvalue $s_{A}$ of $A$; if $s_{A} \notin \sigma(T_{A})$, $P^{(A)}_{s_{A}} = 0$).
\end{itemize}
\end{defn}
The definition of noncontextuality in the CH approach goes as follows \cite{amaral2018graph}.
\begin{defn}[Noncontextuality]\label{Def: Noncontextual}
    A behavior $p$ in a scenario $\mathcal{S}$ is said to be non-contextual if there is a probability distribution $\overline{p}:O^{\mathcal{A}} \ri [0,1]$ satisfying, for any context $C$,
    \[p(\cdot | C) = \overline{p}_{C},\]
    where $\overline{p}_{C}$ denotes the marginal of $\overline{p}$ in $O^{C}$, i.e., for any $s \in O^{C}$,
    \[\overline{p}_{C}(s) \doteq \sum_{\begin{subarray}{l} t \in O^{\mathcal{A}} \\ t|_{C} = s\end{subarray} }\overline{p}(t).\]
\end{defn}
One can easily prove that non-contextuality and classicality (i.e., having a classical realization) are equivalent concepts in the compatibility-hypergraph approach \cite{amaral2018graph}, by which we mean that a behavior $p$ is noncontextual (definition \ref{Def: Noncontextual}) if and only if it has a classical realization (definition \ref{def: Classical behavior}).

It is easy to show that any classical behavior (i.e., any behavior satisfying definition \ref{def: Classical behavior}) has a quantum realization, and it is also easy to show that any quantum behavior is non-disturbing \cite{amaral2018graph}. Therefore, if we denote by $\mathscr{NC}(\mathcal{S})$, $\mathscr{Q}(\mathcal{S})$ and $\mathscr{ND}(\mathcal{S})$ the sets of noncontextual (or, equivalently, ``classical''), quantum and non-disturbing behaviors, respectively, on a scenario $\mathcal{S}$, we obtain the well-known chain of inclusions
\begin{align}
    \mathscr{NC}(\mathcal{S}) \subset  \mathscr{Q}(\mathcal{S}) \subset\mathscr{ND}(\mathcal{S}).
\end{align}
The behavior we are interested in (definition \ref{def: n-faced coin toss}) is nondisturbing but lies outside the quantum set \cite{amaral2018graph}.

\bibliographystyle{plainurl}
\bibliography{Bibliography}

\begin{thebibliography}{10}

\bibitem{Abramsky_2011}
S.~Abramsky and A.~Brandenburger.
\newblock The sheaf-theoretic structure of non-locality and contextuality.
\newblock {\em New Journal of Physics}, 13(11):113036, nov 2011.
\newblock \href {https://doi.org/10.1088/1367-2630/13/11/113036} {\path{doi:10.1088/1367-2630/13/11/113036}}.

\bibitem{AmaralWirings}
B.~Amaral, A.~Cabello, M.T. Cunha, and L.~Aolita.
\newblock Noncontextual wirings.
\newblock {\em Phys. Rev. Lett.}, 120:130403, Mar 2018.
\newblock \href {https://doi.org/10.1103/PhysRevLett.120.130403} {\path{doi:10.1103/PhysRevLett.120.130403}}.

\bibitem{amaral2018graph}
B.~Amaral and M.T. Cunha.
\newblock {\em On Graph Approaches to Contextuality and their Role in Quantum Theory}.
\newblock SpringerBriefs in Mathematics. Springer International Publishing, 2018.
\newblock \href {https://doi.org/10.1007/978-3-319-93827-1} {\path{doi:10.1007/978-3-319-93827-1}}.

\bibitem{AraujoCycle}
Mateus Ara\'ujo, Marco~T\'ulio Quintino, Costantino Budroni, Marcelo~Terra Cunha, and Ad\'an Cabello.
\newblock All noncontextuality inequalities for the $n$-cycle scenario.
\newblock {\em Phys. Rev. A}, 88:022118, Aug 2013.
\newblock URL: \url{https://link.aps.org/doi/10.1103/PhysRevA.88.022118}, \href {https://doi.org/10.1103/PhysRevA.88.022118} {\path{doi:10.1103/PhysRevA.88.022118}}.

\bibitem{armstrong1993world}
David~M Armstrong.
\newblock A world of states of affairs.
\newblock {\em Philosophical Perspectives}, 7:429--440, 1993.

\bibitem{Atmanspacher2019}
Harald Atmanspacher and Thomas Filk.
\newblock {\em Contextuality Revisited: Signaling May Differ From Communicating}, pages 117--127.
\newblock Springer International Publishing, Cham, 2019.
\newblock \href {https://doi.org/10.1007/978-3-030-21908-6_10} {\path{doi:10.1007/978-3-030-21908-6_10}}.

\bibitem{barbosa2021closing}
R.S. Barbosa, M.~Karvonen, and S.~Mansfield.
\newblock Closing bell: Boxing black box simulations in the resource theory of contextuality, 2021.
\newblock \href {https://arxiv.org/abs/2104.11241} {\path{arXiv:2104.11241}}.

\bibitem{BrunnerBell}
Nicolas Brunner, Daniel Cavalcanti, Stefano Pironio, Valerio Scarani, and Stephanie Wehner.
\newblock Bell nonlocality.
\newblock {\em Rev. Mod. Phys.}, 86:419--478, Apr 2014.
\newblock URL: \url{https://link.aps.org/doi/10.1103/RevModPhys.86.419}, \href {https://doi.org/10.1103/RevModPhys.86.419} {\path{doi:10.1103/RevModPhys.86.419}}.

\bibitem{budroni2021quantum}
C.~Budroni, Cabello.A, O.~Gühne, M.~Kleinmann, and JA. Larsson.
\newblock Quantum contextuality, 2021.
\newblock \href {https://arxiv.org/abs/2102.13036} {\path{arXiv:2102.13036}}.

\bibitem{CSW}
A.~Cabello, S.~Severini, and A.~Winter.
\newblock Graph-theoretic approach to quantum correlations.
\newblock {\em Phys. Rev. Lett.}, 112:040401, Jan 2014.
\newblock \href {https://doi.org/10.1103/PhysRevLett.112.040401} {\path{doi:10.1103/PhysRevLett.112.040401}}.

\bibitem{cabello2019simple}
Ad\'an Cabello.
\newblock Quantum correlations from simple assumptions.
\newblock {\em Phys. Rev. A}, 100:032120, Sep 2019.
\newblock URL: \url{https://link.aps.org/doi/10.1103/PhysRevA.100.032120}, \href {https://doi.org/10.1103/PhysRevA.100.032120} {\path{doi:10.1103/PhysRevA.100.032120}}.

\bibitem{cabello2021converting}
Ad\'an Cabello.
\newblock Converting contextuality into nonlocality.
\newblock {\em Phys. Rev. Lett.}, 127:070401, Aug 2021.
\newblock URL: \url{https://link.aps.org/doi/10.1103/PhysRevLett.127.070401}, \href {https://doi.org/10.1103/PhysRevLett.127.070401} {\path{doi:10.1103/PhysRevLett.127.070401}}.

\bibitem{cervantes2018snow}
V{\'\i}ctor~H Cervantes and Ehtibar~N Dzhafarov.
\newblock Snow queen is evil and beautiful: Experimental evidence for probabilistic contextuality in human choices.
\newblock {\em Decision}, 5(3):193, 2018.

\bibitem{deRondeOmelette}
C.~de~Ronde.
\newblock Unscrambling the omelette of quantum contextuality (part i): Preexistent properties or measurement outcomes?
\newblock {\em Foundations of Science}, 25, 03 2020.
\newblock \href {https://doi.org/10.1007/s10699-019-09578-8} {\path{doi:10.1007/s10699-019-09578-8}}.

\bibitem{doring2005kochen}
A.~D{\"o}ring.
\newblock Kochen--specker theorem for von neumann algebras.
\newblock {\em International Journal of Theoretical Physics}, 44(2):139--160, 2005.

\bibitem{WhatThing}
A.~D{\"o}ring and C.~Isham.
\newblock {\em ``What is a Thing?'': Topos Theory in the Foundations of Physics}, pages 753--937.
\newblock Springer Berlin Heidelberg, Berlin, Heidelberg, 2011.
\newblock \href {https://doi.org/10.1007/978-3-642-12821-9_13} {\path{doi:10.1007/978-3-642-12821-9_13}}.

\bibitem{CbD2.0}
E.~Dzhafarov and J.~Kujala.
\newblock Contextuality-by-default 2.0: Systems with binary random variables.
\newblock In Jose~Acacio de~Barros, Bob Coecke, and Emmanuel Pothos, editors, {\em Quantum Interaction}, pages 16--32, Cham, 2017. Springer International Publishing.

\bibitem{Gupta2022}
Shashank Gupta, Debashis Saha, Zhen-Peng Xu, Adán Cabello, and A.~S. Majumdar.
\newblock Quantum contextuality provides communication complexity advantage, 2022.
\newblock URL: \url{https://arxiv.org/abs/2205.03308}, \href {https://doi.org/10.48550/ARXIV.2205.03308} {\path{doi:10.48550/ARXIV.2205.03308}}.

\bibitem{halvorson2019realist}
Hans Halvorson.
\newblock To be a realist about quantum theory.
\newblock In {\em Quantum Worlds: Perspectives on the Ontology of Quantum Mechanics}. Cambridge University Press, 2019.

\bibitem{sep-kochen-specker}
Carsten Held.
\newblock {The Kochen-Specker Theorem}.
\newblock In Edward~N. Zalta, editor, {\em The {Stanford} Encyclopedia of Philosophy}. Metaphysics Research Lab, Stanford University, {S}pring 2018 edition, 2018.

\bibitem{Hofer2021Causal}
Gábor Hofer-Szabó.
\newblock Causal contextuality and contextuality-by-default are different concepts.
\newblock {\em Journal of Mathematical Psychology}, 104:102590, 2021.
\newblock URL: \url{https://www.sciencedirect.com/science/article/pii/S0022249621000687}, \href {https://doi.org/10.1016/j.jmp.2021.102590} {\path{doi:10.1016/j.jmp.2021.102590}}.

\bibitem{howard2014contextuality}
Mark Howard, Joel Wallman, Victor Veitch, and Joseph Emerson.
\newblock Contextuality supplies the ‘magic’for quantum computation.
\newblock {\em Nature}, 510(7505):351--355, 2014.

\bibitem{Matt2019}
Matt Jones.
\newblock Relating causal and probabilistic approaches to contextuality.
\newblock {\em Philosophical Transactions of the Royal Society A: Mathematical, Physical and Engineering Sciences}, 377(2157):20190133, 2019.
\newblock URL: \url{https://royalsocietypublishing.org/doi/abs/10.1098/rsta.2019.0133}, \href {https://arxiv.org/abs/https://royalsocietypublishing.org/doi/pdf/10.1098/rsta.2019.0133} {\path{arXiv:https://royalsocietypublishing.org/doi/pdf/10.1098/rsta.2019.0133}}, \href {https://doi.org/10.1098/rsta.2019.0133} {\path{doi:10.1098/rsta.2019.0133}}.

\bibitem{KCBSeminal}
Alexander~A. Klyachko, M.~Ali Can, Sinem Binicio\ifmmode~\breve{g}\else \u{g}\fi{}lu, and Alexander~S. Shumovsky.
\newblock Simple test for hidden variables in spin-1 systems.
\newblock {\em Phys. Rev. Lett.}, 101:020403, Jul 2008.
\newblock URL: \url{https://link.aps.org/doi/10.1103/PhysRevLett.101.020403}, \href {https://doi.org/10.1103/PhysRevLett.101.020403} {\path{doi:10.1103/PhysRevLett.101.020403}}.

\bibitem{Kochen1975}
S.~Kochen and E.P. Specker.
\newblock The problem of hidden variables in quantum mechanics.
\newblock {\em Journal of Mathematics and Mechanics}, 17(1):59--87, 1967.
\newblock URL: \url{http://www.jstor.org/stable/24902153}.

\bibitem{MarianRule}
Marian Kupczynski.
\newblock Contextuality-by-default description of bell tests: Contextuality as the rule and not as an exception.
\newblock {\em Entropy}, 23(9), 2021.
\newblock URL: \url{https://www.mdpi.com/1099-4300/23/9/1104}, \href {https://doi.org/10.3390/e23091104} {\path{doi:10.3390/e23091104}}.

\bibitem{LeiferIs}
Matthew Leifer.
\newblock Is the quantum state real? an extended review of $\psi$-ontology theorems.
\newblock {\em Quanta}, 3(1):67--155, 2014.
\newblock URL: \url{http://quanta.ws/ojs/index.php/quanta/article/view/22}, \href {https://doi.org/10.12743/quanta.v3i1.22} {\path{doi:10.12743/quanta.v3i1.22}}.

\bibitem{li2017experimental}
T.~Li, Q.~Zeng, X.~Song, and X.~Zhang.
\newblock Experimental contextuality in classical light.
\newblock {\em Scientific reports}, 7(1):1--8, 2017.
\newblock \href {https://doi.org/10.1038/srep44467} {\path{doi:10.1038/srep44467}}.

\bibitem{li2019state}
T.~Li, Q.~Zeng, X.~Zhang, T.~Chen, and X.~Zhang.
\newblock State-independent contextuality in classical light.
\newblock {\em Scientific reports}, 9(1):1--12, 2019.
\newblock \href {https://doi.org/10.1038/s41598-019-51250-5} {\path{doi:10.1038/s41598-019-51250-5}}.

\bibitem{NatureMazurek}
M.D. Mazurek, M.~F. Pusey, R.~Kunjwal, K.~J. Resch, and R.~W. Spekkens.
\newblock An experimental test of noncontextuality without unphysical idealizations.
\newblock {\em Nature communications}, 7, June 2016.
\newblock \href {https://doi.org/10.1038/ncomms11780} {\path{doi:10.1038/ncomms11780}}.

\bibitem{NaturePopescu}
S.~Popescu.
\newblock Nonlocality beyond quantum mechanics.
\newblock {\em Nature Physics}, 10:040403, Apr. 2014.
\newblock \href {https://doi.org/10.1038/nphys2916} {\path{doi:10.1038/nphys2916}}.

\bibitem{Leo}
L.~Santos and B.~Amaral.
\newblock Conditions for logical contextuality and nonlocality.
\newblock {\em Phys. Rev. A}, 104:022201, Aug 2021.
\newblock \href {https://doi.org/10.1103/PhysRevA.104.022201} {\path{doi:10.1103/PhysRevA.104.022201}}.

\bibitem{SpekkensSeminal}
R.~W. Spekkens.
\newblock Contextuality for preparations, transformations, and unsharp measurements.
\newblock {\em Phys. Rev. A}, 71:052108, May 2005.
\newblock \href {https://doi.org/10.1103/PhysRevA.71.052108} {\path{doi:10.1103/PhysRevA.71.052108}}.

\bibitem{Spekkens_2007}
R.~W. Spekkens.
\newblock Evidence for the epistemic view of quantum states: A toy theory.
\newblock {\em Physical Review A}, 75(3), Mar 2007.
\newblock \href {https://doi.org/10.1103/physreva.75.032110} {\path{doi:10.1103/physreva.75.032110}}.

\bibitem{SvozilClicks}
Karl Svozil.
\newblock What is so special about quantum clicks?
\newblock {\em Entropy}, 22(6), 2020.
\newblock URL: \url{https://www.mdpi.com/1099-4300/22/6/602}, \href {https://doi.org/10.3390/e22060602} {\path{doi:10.3390/e22060602}}.

\bibitem{sep-states-of-affairs}
Mark Textor.
\newblock {States of Affairs}.
\newblock In Edward~N. Zalta, editor, {\em The {Stanford} Encyclopedia of Philosophy}. Metaphysics Research Lab, Stanford University, {S}ummer 2021 edition, 2021.

\bibitem{tezzin2023stepbystep}
Alisson Tezzin.
\newblock Step-by-step derivation of the algebraic structure of quantum mechanics (or from nondisturbing to quantum correlations by connecting incompatible observables), 2023.
\newblock \href {https://arxiv.org/abs/2303.04847} {\path{arXiv:2303.04847}}.

\bibitem{wang2021analysing}
Daphne Wang, Mehrnoosh Sadrzadeh, Samson Abramsky, H~V{\'\i}ctor, and VH~Cervantes.
\newblock Analysing ambiguous nouns and verbs with quantum contextuality tools.
\newblock {\em Journal of Cognitive Science}, 22(3):391--420, 2021.

\bibitem{ZhangExperimental}
Aonan Zhang, Huichao Xu, Jie Xie, Han Zhang, Brian~J. Smith, M.~S. Kim, and Lijian Zhang.
\newblock Experimental test of contextuality in quantum and classical systems.
\newblock {\em Phys. Rev. Lett.}, 122:080401, Feb 2019.
\newblock URL: \url{https://link.aps.org/doi/10.1103/PhysRevLett.122.080401}, \href {https://doi.org/10.1103/PhysRevLett.122.080401} {\path{doi:10.1103/PhysRevLett.122.080401}}.

\end{thebibliography}

\end{document}